\documentclass{aa}  
\usepackage{graphicx}
\usepackage{txfonts}
\usepackage[]{natbib}

\def\minspt{$\buildrel{\prime}\over .$}

\def\magspt{$\buildrel{\rm m}\over .$}
\def\simlt{\lower.5ex\hbox{$\; \buildrel < \over \sim \;$}}
\def\simgt{\lower.5ex\hbox{$\; \buildrel > \over \sim \;$}}

\def\3H{{\textstyle{3\over2}}}

\def\HI{H{\small I}}

\def\simgt{\lower.5ex\hbox{$\; \buildrel > \over \sim \;$}}

\begin{document}
   \title{Truncations of stellar disks and warps of HI-layers in 
edge-on spiral galaxies}

\titlerunning{Truncations and warps in edge-on galaxies}  

   \author{P.C. van der Kruit
         \inst{}
          }

   \offprints{P.C. van der Kruit}

   \institute{Kapteyn Astronomical Institute, University of Groningen,
P.O. Box 800, 9700 AV Groningen, the Netherlands\\
              \email{vdkruit@astro.rug.nl}
             }

   \date{Received December 2006, accepted 2007}

 
  \abstract
{Edge-on spiral galaxies often have stellar disks with relatively sharp 
truncations and there is
extensive observational material of warped \HI-layers in the
outer parts. Warps appear to start preferentially near the optical
boundaries of the disks, but the possible relation between truncations 
and warps has not been studied in detail. }
{The aim is to make a comparative study of warps and truncations in 
edge-on galaxies. Samples with detailed surface photometry or
\HI-mapping have little overlap. The Garc\'{\i}a-Ruiz et al. (2002) sample 
with extensive Westerbork \HI-mapping is a candidate to be complemented
with luminosity distributions from other sources.}
{The Sloan Digital Sky Survey (and in a few cases the STScI Digital Sky
Survey) has been used to provide these data. The method to identify truncations in these
digital datasets has been tested using the sample of edge-on galaxies of
van der Kruit \&\ Searle (1981a/b, 1982a/b).} 
{(i.) The majority (17 out of 23)
of the galaxies show evidence for
truncations, consistent with previous findings in other samples.
(ii.) When an \HI-warp is present it starts at 1.1 truncation radii 
(statistically allowing all possible geometries).
(iii.) This supplements the rules for warps formulated by Briggs
(1990), if the Holmberg radius is replaced for edge-on systems with the 
truncation radius.
(iv.) The truncation radius and the onset of the warps coincide radially
sometimes with features in the rotation curve and often with steep
declines in the \HI\ surface density. The latter is also true for
less inclined systems.
(v.) Inner disks are very flat and the onset of the warp just beyond the
truncation radius is abrupt and discontinuous.
(vi.) In an appendix the definition and derivation of
the Holmberg radius is discussed. Contrary to what is often assumed in
the literature Holmberg radii are not corrected for inclination.}
{These findings suggest that the inner flat disk and the outer warped
disk are distinct components with quite different formation histories,
probably involving quite different epochs. The inner disk
forms initially (either in a monolithic process in a short period or
hierarchically on a somewhat
more protracted timescale) and the warped outer disk
forms as a result of much later infall of gas with a higher angular
momentum in a different orientation. The results are also consistent
in this picture with an origin of the disk truncations that is related
to the maximum specific angular momentum available during its formation.}

   \maketitle
%

\section{Introduction}

It has been known for a long time that the neutral hydrogen (\HI) 
layer of our Galaxy is warped in the outer parts. It was discovered
independently in early surveys of the Galactic \HI\ in the north by
Burke (1957) and in the south by Kerr and Hindman (Kerr, 1957). In
external spiral galaxies the first indication came from the work of 
Rogstad et al. (1974), when they obtained aperture synthesis 
observations of the \HI\ in M83. Both the distribution and the velocity field 
of the \HI\ showed features that could be interpreted as a warping of
the gaseous disk in circular motions in inclined rings. This later
became known as the `tilted-ring' description for `kinematic
warps' and many more galaxies have been shown to have such deviations 
using this method (e.g. Bosma, 1981). The case that this warping occurs
in many spiral galaxies was strengthened by the early observations of
edge-on systems. Sancisi (1976) was the first to perform such
observations and showed that the \HI\ in four out of five observed edge-ons 
(among which NGC 4565 and NGC 5907)  displayed strong deviations
from a single plane. Sancisi (1983) discussed these warps in somewhat
more detail and in particular noted that in the radial direction the \HI\ 
surface densities often display steep drop-offs followed by a ``shoulder'' or
``tail'' at larger radii.

The most extreme (`prodigious') warp in an edge-on system was observed by
Bottema et al. (1987; see also Bottema 1995, 1996) in NGC 4013. Recently 
Garc\'{\i}a-Ruiz et al. (2002; also see
Garc\'{\i}a-Ruiz, 2001) presented 21-cm observations of a sample of 26 
edge-on galaxies in the northern hemisphere. This showed that \HI-warps
are ubiquitous; in fact state that {\it ``all galaxies
that have an extended \HI-disk with respect to the optical are
warped''}. 
Studies of possible warps in the stellar
disks have also been made (for recent results see e.g. de Grijs, 1997, 
ch. 9 and Reshetnikov et al. 2002); although there is evidence for
such warps in most edge-on galaxies the amplitude is very small compared
to what is observed in the \HI.

The origin of warps has been subject of extensive study and has been
reviewed recently in for example Garc\'{\i}a-Ruiz et al. (2002), Shen
\&\ Sellwood (2006) and Binney (2006). Although none of the models is
completely satisfactory, most workers seem to agree that it must have
something to do with a constant accretion of material with an angular
momentum vector that has another orientation than the main disk. In
models bij Jiang \&\ Binney (1999) and Shen \&\ Sellwood (2006) this
results in an inclined outer torus in the dark halo
 that distorts the existing disk and
causes it to become warped. The posssibility of a misalignment in the
angular momenta and therefore the principal planes between
the stellar disk and the dark halo (Debattista \&\ Sellwood, 1999) has
recently received some observational support from the observations of
Battaglia et al. (2006) of NGC 5055. The \HI\ data suggest that the
inner flat disk and the outer warped part of the \HI\ have different
kinematical centers and systemic velocities, suggesting a dark halo
with not only a different orientation, but also an offset with respect
to the disk.

Another aspect of disks in galaxies is the fact that the {\it stellar} disks 
appear to have `truncations'. This was first noticed by van der Kruit (1979) 
in photographic surface photometry of NGC 4244, 4565 and 5907. Two 
of these have no bulges and `vertical' profiles showed that stellar
disks have roughly constant thickness with galactocentric radius. Van 
der Kruit and Searle (1981a,b, 1982a,b) extended
this to a sample of 7 galaxies and performed a much more detailed
analysis of the surface brightness distributions. It was found that
these relatively sharp outer edges correspond to a rather sudden
decrease in the radial e-folding of the surface brightness distributions, 
dropping to values below 1 kpc. The truncation
radius $R_{max}$ usually occurs around 4 to 5 radial scalelengths $h$.

This early surface photometry was obtained from photographic plates. The
existence of the truncations --and the values for scale parameters 
(scalelength and scaleheight) of these stellar disks-- and the associated
truncation radii were confirmed to good or exellent precision using
CCD photometry by Morrison et al. (1994) for NGC 5907, by Fry et al. (1999)
for NGC 4244, by Morrison et al. (1997) and Xilouris et al. (1998) for NGC 891
and by Wu et al. (2002) for NGC 4565.

The existence of truncations has been subject of extensive study, both in
edge-on galaxies (e.g. Kregel \&\ van der Kruit, 2004) and in disks at
various orientations (Pohlen \&\ Trujillo, 2006; Trujillo \&\ Pohlen, 2006;
Florido et al., 2006)). 
In the edge-on galaxies the signature is quite clear; the radial light profile
simply drops well below an extrapolation of the radial light profile of 
the inner parts. Kregel,
van der Kruit \&\ de Grijs (2002) conclude that in their sample of 34 southern
edge-ons (defined by de Grijs, 1997)
evidence for truncation is seen in at least 60\%. In less inclined
systems (e.g. Pohlen \&\ Trujillo, 2006) also a frequency of around 60\%\ 
has been found, which shows that independent of orientation
the conclusion can be drawn that truncations are common among galaxies.

In the less inclined systems studies referred to above have found 
a diversity of behaviour, ranging from a rather
sharp truncation through a continuation into the noise of the
exponential profile to an actual flattening. In the edge-on systems the
truncation is seen best as a lack of growing diameter with fainter light
levels. In fact, that is how it was first noted in NGC 4565 (van der
Kruit, 1979). Freeman (2006) has reviewed
the subject recently and documents that although truncations as in NGC
4565 are common, some (not edge-on) galaxies without truncations do exist,
e.g. NGC 300 (Bland-Hawthorn et al. 2005). 

\begin{figure*}
\centering
\includegraphics[width=18cm]{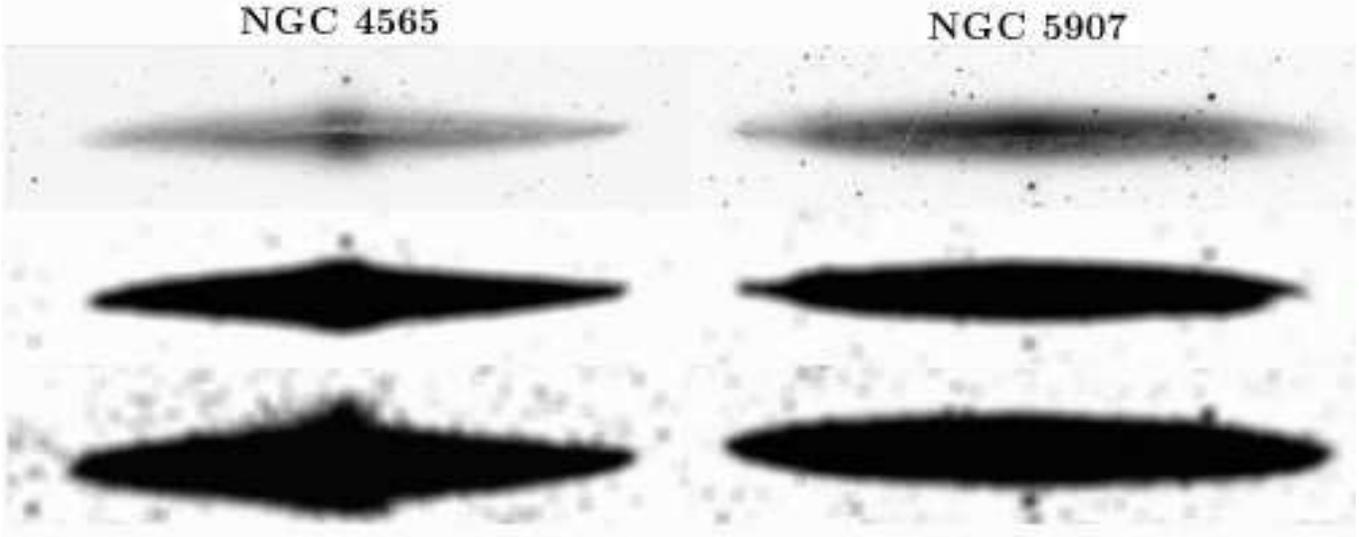}
\caption{Images of NGC 4565 and NGC 5907 
from the Sloan Digital Sky Survey. The upper ones are the grey-scale version
of the original picture without any change in the look-up table. 
The lower ones have been clipped at a shallow and a deep
level (for more precise information see the text) and subsequently
have been heavily smoothed. Although the images grow 
significantly toward fainter
levels in the vertical direction, the change in the radial direction is
much smaller.}
\label{Fig1}
\end{figure*}

The origin of the truncations is still not clear. The early suggestion in 
van der Kruit \&\ Searle (1981a) that it may be related to the disk 
formation model of Larson (1976) is probably no longer relevant, since
that scenario of a very slowly growing disk does not agree with the fact that
even in the outer parts the ages of the stellar disks appear several
Gyrs at least (de Jong, 1996; Bell \&\ de Jong, 2000). Although it was
speculated in van der Kruit \&\ Searle (1981a) that the truncations may
correspond to points where the gas density drops below a value
critical for star formation, Fall \&\
Efstathiou (1980) suggested that it was the shear of differential
rotation that inhibits star formation when it becomes large enough to 
overcome self-gravity in the gas. Van der Kruit \&\ Searle (1982a) investigated
that by estimating the numerical value of parameters in the Goldreich
\&\ Lynden-Bell (1965) criterion for gravitational stability in a
rotating gaseous disk. This showed that the criterion systematically
underpredicts the value of $R_{max}$ by 20\%\ or so, but this does not
prove that it is unrelated to this criterion. In fact Schaye (2004) has
studied the possibility of a threshold in the gas density for star
formation to occur as the cause of the truncations. The third
possibility is that it results from a well-defined maximum in specific
angular momentum in the protogalaxy (van der Kruit, 1987) 
if the collapse occurs with detailed
conservation of specific angular momentum (Fall \&\ Efstathiou, 1980). 

Kregel \&\ van der Kruit (2004) have discussed this issue in the light of
the observations in the de Grijs sample. The ratio $R_{max}/h$ decreases
towards larger scalelengths and there is a correlation of this
ratio with face-on central surface brightness, larger ratio's
corresponding to brighter galaxies. This makes the disk surface
brightness at the truncation radius rather constant.  
Studies of less inclined systems, as in Pohlen \&\ Trujillo 
2006), have come to the similar correlations, confirming that these
conclusions are also independent of orientation. These obervations
do most readily fit into the model of a star-formation
threshold; for it to fit into the maximum specific angular momentum scenario
some redistribution of angular momentum might be required. On the
other hand, as I will document more fully in the discussion below, features
in rotation curves of at least some edge-on galaxies indicate that the 
truncations correspond to sharp drops in the total matter density in the 
disks as well, which would favor the maximum specific angular momentum
cause of the truncations.

The earliest observations of \HI-warps suggested that these warps start
at the edges of the optical disks. It certainly was very obvious in the
warp for the edge-on galaxies NGC 5907 and NGC 4013. It was suggested by
Bottema et al. (1987) that the coincidence of the onset of the warp in
NGC 4013 with the truncation radius might be a factor in explaining the
persistence of warps. Briggs (1990) used existing observations and
tilted-ring models in moderately inclined galaxies to define a set of
``rules of behavior for galactic warps''. One was that ``warps
change character at a transition radius near $R_{Ho}$''. The latter
radius is the so-called Holmberg-radius listed for 300 bright galaxies in
Holmberg (1958); it corresponds in a statistical sense
to the isophote of 26.0 photovisual
mag arcsec$^{-2}$ or 26.5 photographic mag arcsec$^{-2}$. In the appendix
I make some observations concerning the use of the Holmberg
radius and its relation to the truncation radius in edge-on systems.

A possible coincidence of the onset of the warp and the truncation
radius of the stellar disk could be important for our understanding of
the origin of warps. It would point to a picture in which the stellar
disk formed in an initial stage (which could last several Gyr),
resulting in a very flat disk with a truncation. Later on then more gas
might fall in, which in general would not have the same orientation for
its angular momentum. Although some of it may settle in the existing
disk, much of it may have high angular momentum 
and settle at larger radii and in different planes. Support for
this picture comes from investigations of features in rotations curves
and in the \HI\ surface densities at the radii of the truncations and the
onset of the warp (van der Kruit, 2001). I will return to this below.

The coincidence of truncation and onset of a warp can be studied by
considering a sample of edge-on galaxies for which both 21-cm line
observations and optical surface photometry are available. Unfortunately
the sample of de Grijs with surface photometry is in the south and the
sample of Garc\'{\i}a-Ruiz in the north. The sample of
Garc\'{\i}a-Ruiz is part of the WHISP\footnote{Westerbork observations of
neutral Hydrogen in Irregular and SPiral galaxies, see
www.astro.rug.nl/~whisp/.} survey 
(Kamphuis, Sijbring \&\ van Albada, 1996) for which surface photometry is
part of the project. However, as explained by Garc\'{i}a-Ruiz (2002), 
{\it ``the
quality of the optical images is not good enough to perform photometry
on them, possiby due to tracking errors and large background gradients}''.
This material is not suitable for determinations of possible truncations.
Garc\'{\i}a-Ruiz (2002) and Garc\'{\i}a-Ruiz et al. (2004) use the $R_{25}$
radius from the LEDA catalogue and find that warps usually start
around the edge of the optical disk. In this paper I will make
that more quantitative and look for evidence of truncations in this
sample in data available in digital sky surveys.

\section{Observing truncation with the use of digital sky surveys}

The purpose of this section is to describe where surface
photometry for the edge-on galaxies in the sample described could be
obtained from existing datasets
and how this material then can been used to evaluate possible
evidence for truncations. The application of this method will be the 
subject of the next section.

\begin{table*}
\caption{Diameters along the major axis, expressed as fractions of twice 
the truncation radius as given by van der Kruit \&\ Searle, 
measured at various clip levels
for 5 prominent edge-on galaxies that have clear evidence for truncations.}
\label{Table1}
\centering
\begin{tabular}{c c c c c c c}
\hline\hline
Cliplevel & \multicolumn{6}{c}{$D_{major axis}/(2\times R_{max}$)} \\
\hline
 & NGC 4013 & NGC 4217 & NGC 4244 & NGC 4565 & NGC 5907 & Mean \\
\hline
245 & 1.03 & 1.03 & 0.98 & 0.99 & 1.02 & 1.01 $\pm $ 0.02 \\
240 & 0.99 & 1.01 & 0.97 & 0.99 & 1.00 & 0.99 $\pm $ 0.01 \\
235 & 0.97 & 1.00 & 0.95 & 0.99 & 0.99 & 0.98 $\pm $ 0.02 \\
230 & 0.95 & 0.98 & 0.92 & 0.98 & 0.99 & 0.96 $\pm $ 0.03 \\
225 & 0.94 & 0.96 & 0.91 & 0.98 & 0.98 & 0.95 $\pm $ 0.03 \\
220 & 0.93 & 0.95 & 0.90 & 0.98 & 0.98 & 0.95 $\pm $ 0.03 \\
215 & 0.93 & 0.94 & 0.89 & 0.96 & 0.97 & 0.94 $\pm $ 0.03 \\
210 & 0.89 & 0.89 & 0.77 & 0.95 & 0.96 & 0.89 $\pm $ 0.08 \\
205 & 0.87 & 0.85 & 0.74 & 0.94 & 0.95 & 0.87 $\pm $ 0.08 \\
200 & 0.83 & 0.83 & 0.69 & 0.92 & 0.91 & 0.84 $\pm $ 0.09 \\
\hline
\end{tabular}
\end{table*}

The prime dataset is the Sloan Digital Sky Survey (SDSS). It covers a
large fraction of the northern sky
and has photometric calibrations accurate to a few tenths of a magnitude.
The work decribed here was performed in May, 2006, using the 
Fourth Data Release (DR4). Images can be obtained at
http://cas.sdss.org/dr4/en/tools/chart/chart.asp. 
This covered 6 of the 8 galaxies in the sample of van der Kruit
\&\ Searle and 20 of the 26 galaxies of the Garc\'{\i}a-Ruiz et al.
sample. DR5 (http://cas.sdss.org/dr5/en/tools/chart/chart.asp)
has since then appeared, and respectively one and two more galaxies have 
become available in the samples from this new release. The ones that remain
missing in the SDSS are NGC 891 in the van der Kruit \&\ Searle
sample and UGC 1281, UGC 2459, UGC 3137 and UGC 3909 
in the Garc\'{\i}a-Ruiz et al. sample. 

First we consider the van der Kruit \&\ Searle sample in the SDSS. In
Fig.~\ref{Fig1} I show two of these galaxies to illustrate
the way these data have been analysed. These are NGC 4565 (in which
actually the evidence for truncations was first noticed) and NGC 5907.
The first has a bulge, but the second is probably almost entirely
disk. The images were obtained throught the ``Finding Chart Tool'',
usually downloading a 1024$^2$ image. This was treated with the `xv'
routine in a Unix/Linux environment. In the case of 
Fig.~\ref{Fig1} the images were rotated so that the
disks appear horizontal. The data format was then changed from 24-bit to
8-bit mode, converted to grey-scale and to inverse video. The upper panels
show the images after this treatment. Next they were clipped at two
different levels. A shallow clip was produced by setting all values
in the image below 215 to 'white' and the rest to 'black'. The deep clip
was produced the same way but now with 245 as the dividing value. These
images were then smoothed ('blurred') with a mask of about
15 pixels. The results are in the lower two panels.

Fig.~\ref{Fig1} clearly shows what the signature of a truncation in
such images. The minor axes do grow
significantly at faint levels, especially in the case of NGC 4565 due
to the relatively prominent bulge. But in the direction of the major
axes the increase in diameter is very small. It therefore
appears that evidence for a truncation can be searched for by
measuring the axis ratios in these images. All images from the SDSS
have been treated in this manner. By measuring the minor axis 
diameters and comparing to the isophote maps in the original papers, I
find that rather consistently (to an accuracy of about 0.3 magnitudes)
the diameters measured correspond roughly to the
isophotes at 26.5 and 27.5 mag arcsec$^{-2}$ in the $J$-Band of van der
Kruit \&\ Searle. This is very close to Johnson $B$-Band.

\begin{table*}
\caption{Dimensions (in arcmin unless ratios) of edge-on galaxies in 
the van der Kruit \&\ Searle sample as measured from the Sloan Digital
Sky Survey and from the Digital Sky Survey 2 (red). The radius $R_{max}$
is the truncation radius found by van der Kruit \&\ Searle and has been
listed for comparison.Values for radii and axis ratios have been rounded
off to one decimal, but their ratios have been calculated before that.}
\label{Table2}      
\centering           
\begin{tabular}{l c c c c c c c c}    
\hline\hline                 
Galaxy & $R_{max}$ & $R_{shallow}$ & $R_{deep}$ & 
\underline{$R_{deep}$} &  
\underline{$R_{deep}\ \ \ $} & $(a/b)_{shallow}$  
& $(a/b)_{deep}$ & \underline{$(a/b)_{deep}\ \ \ $} \\ 
 & ($\prime$) & ($\prime$)  & ($\prime$) & $R_{max}\ \ \ $ &  
$R_{shallow}$ & 
&  & $(a/b)_{shallow}$ \\    
\hline                     
\multicolumn{9}{c}{SDSS} \\
\hline
NGC 4013 & 2.7 & 2.5 & 2.8 & 1.03 & 1.10 & 3.2 & 2.3 & 1.41 \\
NGC 4217 & 3.3 & 3.1 & 3.4 & 1.03 & 1.08 & 3.4 & 2.8 & 1.22 \\
NGC 4244 & 9.5 & 8.5 & 9.3 & 0.98 & 1.09 & 8.0 & 7.0 & 1.15 \\
NGC 4565 & 8.3 & 8.1 & 8.2 & 0.99 & 1.02 & 7.3 & 6.0 & 1.20 \\
NGC 5023 & 3.3 & 2.8 & 3.3 & 1.00 & 1.19 & 6.2 & 5.8 & 1.07 \\
NGC 5907 & 6.0 & 5.8 & 6.1 & 1.02 & 1.06 & 7.3 & 6.3 & 1.16 \\
NGC 7814 & 4.0 & 3.6 & 4.3 & 1.07 & 1.18 & 2.1 & 2.0 & 1.03 \\ 
\hline                                  
\multicolumn{9}{c}{DSS2 (red)} \\
\hline
NGC \ \ 891 & 7.5 & 7.0 & 7.4 & 0.99 & 1.07 & 5.8 & 3.4 & 1.49 \\
NGC 4013 & 2.7 & 2.5 & 2.7 & 1.00 & 1.10 & 3.5 & 2.9 & 1.23 \\
NGC 4217 & 3.3 & 3.2 & 3.3 & 1.00 & 1.04 & 3.4 & 3.0 & 1.15 \\
NGC 4244 & 9.5 & 8.4 & 8.7 & 0.92 & 1.03 & 7.3 & 6.8 & 1.06 \\
NGC 4565 & 8.3 & 8.0 & 8.2 & 0.99 & 1.04 & 7.1 & 5.9 & 1.21 \\
NGC 5023 & 3.3 & 2.5 & 3.2 & 0.97 & 1.30 & 5.9 & 6.3 & 0.95 \\
NGC 5907 & 6.0 & 5.7 & 6.1 & 1.02 & 1.07 & 8.1 & 7.4 & 1.10 \\
NGC 7814 & 4.0 & 2.7 & 3.8 & 0.95 & 1.44 & 2.4 & 2.3 & 1.05 \\
\hline
\end{tabular}
\end{table*}

Images were retrieved then from the SDSS of size 1024$^2$ 
such that the galaxy extended over half to three-fourth of the image.
These were then treated as described above and subsequently
printed on a uniform scale.
The diameters along both the major and minor axes were measured from these
prints.
For all galaxies the actual minor axis was measured and no correction for the
presence of a bulge was attempted. The effect of the bulges is being
addressed below when the results are dicussed.
Independent repeats showed that these measurements can easily be 
done with very good consistency.
The scales of the prints were derived by measuring distances between stars
that could be identified in the contour plots in the van der Kruit \&\ 
Searle publications. I will now consider the choice of the two clip
levels of 215 and 245.

In Table~\ref{Table1} I show the diameters (compared to twice the
truncation radii) for 5 of the galaxies in the sample at various clip
levels. The value 250 was too close to the sky level. One can see that
the average values roughly show the same pattern to about the value
of 215, but then at brighter leves (lower numerical values for the clip)
the deviations between the galaxies used here
suddenly increase. It seems then that the range 215--245 is a useful
range for our purposes. It should also be noted that at the level around 250, 
which is very close to the sky level, one can judge how well the sky
has been determined. This is especially important since for galaxies with
large angular size the data are usually not taken on a single chip or during a
single night. By looking at faint levels one can also judge the continuity
of the isophotes in these clipped images. In none of the galaxies (nor 
for the ones in the sample described below) any evidence was found that this
could be a problem, even though in a few cases the effects could be
seen in the corners of the image. It can also be checked in principle
by examining the two sides of some galaxies and actually measuring radii rather
than diameters; this means that one has to adopt a position of the
center, or at least a position consistent between the two clips. 
It was performed as a trial on NGC 4565 and NGC 5907, but no effect was found.

The results of the measurements are presented in the upper part of
Table~\ref{Table1}. $R_{max}$ is the truncation radius
obtained by van der Kruit \&\ Searle\footnote{For NGC 4013 the pixelsize 
quoted in van der Kruit \&\ Searle
(1982a) was incorrect by a factor 1.5, as discoverd by Bottema (1995).
This correction has been applied.}.  Note that  
diameters were measured rather then radii, which were then divided by two 
in order to compare to $R_{max}$.

Now I turn to Table~\ref{Table2}.
First look at the ratio of the major axis in the deep clip to
truncation radii of van der Kruit \&\ Searle: the mean ratio is 1.02
$\pm 0.03$, showing that indeed the deep clip is a measure for $R_{max}$. 
Next I turn tot the ratio of the dimensions
as measured on the two clips. 
Except for two cases (NGC 5023 and NGC 7814) this is less than 
1.10 and all these galaxies show clear evidence
for a truncation in the detailed photometry of van der Kruit \&\ Searle. 
NGC 5023 is a pure disk galaxy and has been shown to
be a late-type dwarf system (see also Bottema et al., 1986). In the van
der Kruit \&\ Searle (1982a) sample the evidence for the truncation is the
weakest of all; in Fig.~8 of that paper we see in the righthand
panel that the averaged radial profile at the last measured point is
only a few tenth of a magnitude below the extrapolated exponential disk.
Although this galaxy still is believed to have a truncation, it may be
no surprise that it is not readily seen with this method. The fact that 
we see little evidence for it here actually strengthens the case for
the applicability of the approach. NGC 7814 is bulge dominated
and no truncation in the surface brightness distribution exists,
although there is some evidence that the faint disk displays one, which 
becomes visible after subtraction of the exyended $R^{1/4}$
spheroid in the surface brightness maps.

We can also look at the ratio of $b/a$ between the two
clips (see Table~\ref{Table2}). 
The value for the disks with a clear truncations is substantially
different from 1.0. In cases of significant bulges (NGC 4013 and NGC 4217)
it is larger than 1.2, for cases of pure disks (NGC 4244 and NGC 5907) it
between 1.1 and 1.2. For NGC 5023 it is smaller than 1.1, 
because the truncation becomes noticable only at very fainter levels
and in NGC 7814 because the light is dominated by the spheroid.

I will on the basis of this then require for evidence for a truncation
that the ratio between the deep and shallow diameters is 1.10 or less. The
ratio between the axis ratios is used as a further check and this ratio 
should in general exceed 1.10. The cases of NGC 5023 and NGC 7814 show that 
disks without evidence for truncation or bulge-dominated galaxies will have 
this latter ratio close to 1.0.

In order to try and do an analysis of the full sample I looked into
the possibility of applying the same method to images from the STScI
Digital Sky Survey, available at 
http://archive.stsci.edu/cgi-bin/dss\_form. For these I used
the ``POSS2/UKSTU Red'' images. Usually an area of
15$^2$ minutes of arc was retrieved at the standard pixel (about 1 arcsec).
In these images the sky background does not always have the same pixel
value. ``Shallow'' clips were produced with the value of the clip
set such that the noise of the background was just
visible and ``deep'' images when the galaxy was only just noticable
against the backgound noise. Comparison with existing photometry
showed that the levels were both of order one magnitude brighter than in
the analysis on the SDSS images. An image of NGC 891 is
now available. 

The bottom part of Table~\ref{Table2} shows that again the
evidence for the presence of a truncation  is a small increase in the
radius between the shallow and deep images, accompanied by a larger
increase in the axis ratio. NGC 4244 does not provide
evidence for the truncation, but from the actual diameters we see that
the outer edge cannot be seen very well on the image. The conclusion is that
these images can be used as well, but since they do not go to 
very faint levels truncations are less readily visible.

\section{Application to the Garc\'{\i}a-Ruiz et al. sample}

\begin{table*}
\caption{Results of the search for truncation radii and comparison to
the results for the ``hunting for warps'' of Garc\'{\i}a-Ruiz et al. (2002). 
The galaxy types have been taken from that paper.
The asterix in the third
column indicates the galaxies for which the DSS2 had to be used. $R_{\HI}$
is the radius at which the face-on \HI\ surface brightness is 1 M$_{\odot }$
pc$^{-2}$ and has been taken from Garc\'{\i}a-Ruiz et al.}
\label{Table3}      
\centering           
\begin{tabular}{l l c c c c c c c c l}    
\hline\hline                 
Galaxy & Type &  & \underline{$R_{deep}\ \ \ $} & 
\underline{$(a/b)_{deep}\ \ \ $} & 
trunc.? & $R_{deep}$ & $R_{\HI}$ &
$R_{warp}$ & \underline{$R_{warp}$} & Remarks \\
 & & & $R_{shallow}$ & $(a/b)_{shallow}$ & 
 & ($\prime$) & ($\prime$) & ($\prime$) & $R_{max}\ \ \ $ & \\
\hline
UGC 1281 & Sc & * & 1.10 & 1.14 & yes & 2.1 & 3.4 & 1.5? & 0.7? & 
Warp very asymmetric \\  
UGC 2459 & Scd & * & 1.07 & 1.17 & yes & 2.4 & 3.9 & 2.5\ \ \ & 1.0\ \ \ & 
Asymmetric, NE side only \\
UGC 3137 & Sbc  & * & 1.06 & 1.34 & yes & 2.5 & 3.5 & 3.0\ \ \ & 1.2\ \ \ &
Symmetric warp \\
UGC 3909 & SBc & * & 1.09 & 1.15 & yes & 1.2 & 2.3 & 1.1\ \ \ & 0.9\ \ \ &
$R_{warp}$ is 1\minspt 3 on E side, 0\minspt 8 on W side \\  
UGC 4278 & SBc & & 1.05 & 1.29 & yes & 2.6 & 3.1 & --\ \ \  & --\ \ \ &
$R_{\HI}/R_{deep} \sim 1.2$ \\
UGC 4806 & Sc & & 1.07 & 1.12 & yes & 2.1 & 2.3 & ?\ \ \  & --\ \ \ &
Warp not clearly defined; $R_{\HI}/R_{deep} \sim 1.1$ \\ 
UGC 5452 & Sc & & 1.09 & 1.18 & yes & 1.6 & 1.9 & 1.8\ \ \ & 1.1\ \ \ &
Warp and \HI\ beyond $R_{deep}$ only on SW side \\  
UGC 5459 & SBc & & 1.07 & 1.17 & yes & 2.7 & 3.6 & 2.5\ \ \ & 0.9\ \ \ &
$R_{warp}$ is 2\minspt 5 on NW side, 2\minspt 6 on SE side \\  
UGC 5986 & SBd & & 1.13 & 1.22 & no  & 2.4 & 5.3 & --\ \ \ & --\ \ \ &
Optical/\HI\ plume on NE side detached?, \\
 & & & & & & & & & & 
companion on SW side. $R_{deep}$ for main body \\  
UGC 6126 & SBcd & & 1.05 & 1.22 & yes & 2.0 & 3.1 & 1.6\ \ \  & 0.8\ \ \ &
Symmetric warp \\
UGC 6283 & Sab & & 1.13 & 1.22 & no & 2.1 & 3.5 & 3.0\ \ \ & --\ \ \ &
Warp sets in beyond $R_{deep}$ \\
UGC 6964 & SBcd & & 1.05 & 1.30 & yes & 2.2 & 2.8 & 1.7\ \ \ & 0.8\ \ \ &
Symmetric warp \\
UGC 7089 & Sc & & 1.26 & 0.97 & no & 2.1 & 2.1 & --\ \ \ & --\ \ \ &
No \HI\ beyond $R_{deep}$ \\
UGC 7090 & SBc & & 1.20 & 1.02 & no & 3.6 & 3.8 & 2.3\ \ \ & --\ \ \ &
Asymmetric, small warp on N side only \\
UGC 7125 & SBd & W & 1.03 & 1.30 & yes & 2.5 & 5.8 & 2.3\ \ \ & 0.9\ \ \ &
Optical and \HI\ asymmetric, warp on W side \\
 & & E & 1,40 & 0.92 & no  & 2.0 & 3.5 & --\ \ \ & --\ \ \ &
No warp on E side \\  
UGC 7151 & SBc & & 1.06 & 1.14 & yes & 3.4 & 3.3 & --\ \ \ & --\ \ \ &
No \HI\ beyond $R_{deep}$ \\
UGC 7321 & Sc & & 1.07 & 1.39 & yes  & 2.9 & 3.3 & 2.5\ \ \  & 0.9\ \ \ &
Somewhat asymmetric \\
UGC 7483 & SBc & & 1.10 & 1.22 & yes & 2.1 & 2.5 & --\ \ \ & --\ \ \ &
$R_{\HI}/R_{deep} \sim 1.2$ \\
UGC 7774 & Sc & & 1.07 & 1.12 & yes  & 1.8 & 2.7 & 2.0\ \ \ & 1.1\ \ \ &
Rather symmetrical warp \\
UGC 8246 & SBc & W & 1.07 & 1.27 & yes & 1.7 & 2.1 & 1.5\ \ \ & 0.9\ \ \ &
Not precisely edge-on?; very asymmetric \\
 & & E & 1.14 & 1.19 & no & 1.8  & 2.4 & --\ \ \ & --\ \ \ &
No warp on E side \\  
UGC 8286 & Sc & & 1.09 & 1.17 & yes & 3.4 & 4.2 & 3.3\ \ \ & 1.0\ \ \ &
Warp asymmetric, bends back on NE side \\
UGC 8396 & SBc & & 1.33 & 0.91 & no & 1.4 & 1.8 & 0.9\ \ \ & --\ \ \ &
Warp only on NW side \\
UGC 8550 & SBc & & 1.30 & 0.96 & no & 2.1 & 2.8 & --\ \ \ & --\ \ \ &
No clear indication of a warp \\
UGC 8709 & SBbc & SE & 1.09 & 1.00 & yes & 3.4 & 3.5 & --\ \ \ & --\ \ \ &
Bright star on NW side; not entirely edge-on  \\
 & & & & & & & & & & 
No \HI\ beyond $R_{deep}$ \\
UGC 8711 & SBc & & 1.08 & 1.19 & yes & 2.4 & 3.1 & 1.7\ \ \ & 0.7 &
Warp symmetric, but very sharp on SE side \\
UGC 9242 & Sc & & 1.22 & 1.05 & no & 3.0 & 3.1 & --\ \ \ & --\ \ \ &
Superthin galaxy; no \HI\ beyond $R_{deep}$  \\
\hline
\end{tabular}
\end{table*}

Images of the galaxies in the Garc\'{\i}a-Ruiz et al. sample were retrieved
from the SDSS if available and for the remaining four (see above) 
from the DSS2. These images were treated exactly as described above. The
results have been collected in Table~\ref{Table3}. In this table I also
note the diameter in \HI\ (at 1 M$_{\odot}$/pc$^2$, $R_{\HI}$) as given 
by Garc\'{\i}a-Ruiz et al. in their Table~4. Galaxies with an asterix in
the third column are the ones for which the DSS2 had to be used.

First I note that there are two galaxies, UGC 7125 and UGC 8246, for
which the two sides on the sky are entirely different both in the
properties of the stellar disks and the \HI-distributions. One
possibility would be to treat each side as a separate galaxy disk,
but I will take the point of view that these are disturbed
systems that are not suitable for inclusion in a statistical discussion
of general properties of ``normal'' systems. These two galaxies will
not further be considered here. I will also not consider UGC 5986. 
The morphology of this system also shows evidence for
disturbances; there is a companion very nearby on the sky on the SW
side. On the NE side there is a plume that is continuous in velocity
with the main body. It could be a companion or otherwise a
disturbed outer region in the disk. In either case, it is not a normal
system and I exclude it for the same reasons as the other two.

The column `trunc.?' indicates whether or not there is evidence for
a truncation according to the criteria adopted above. 
In that case I assume that the truncation radius $R_{max}$
is equal to $R_{deep}$. It is interesting to note that 
Garc\'{\i}a-Ruiz et al. estimated the optical radius from their images by eye
or took them from NED; 
for galaxies with a truncation their radius equals 1.01 $\pm$\ 0.12 times the
truncation radius determined here.
For three galaxies --UGC 1281, UGC 4806 and UGC 7774-- 
the major axis ratio is less than 1.10, which points at a
truncation, while the ratio of the apparent axis ratios is 
only slightly larger than 1.10. Although this is in agreement with my 
criteria, one might question 
how strong the evidence for a truncation really is for these systems. 
I note that in the van der Kruit \&\ Searle sample NGC 5907 is also such a
case (see Table~\ref{Table2}) and I
will treat these as having truncations on the basis of the
fact that the major axis ratio between the two clips is less than 1.1.
With these included we then find that the method applied here has
provided evidence for truncations in 17 out of 23 galaxies. This is a
significant fraction, similar to the conclusion reached in 
Kregel et al. (2002). The range in Hubble types for this sample is
very small (by design it consists of late types due to the requirement 
to contain observable amounts of \HI) and therefore a dependence of the
occurence of truncation upon Hubble type is not possible; the only early 
type UGC 6283 shows no evidence for a truncation.

\begin{figure}
\centering
\includegraphics[width=8.8cm]{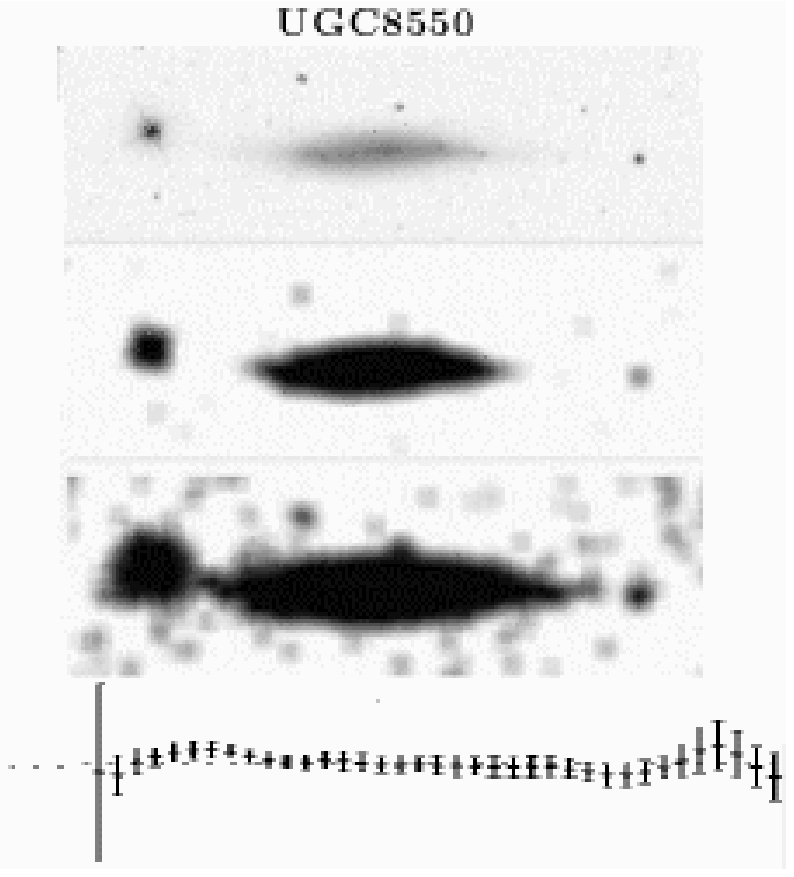}
\caption{UGC 8550 shows no evidence for a truncation nor for a warped 
\HI-layer. The \HI\ is more extended than the optical disk. At the top the
three images from the SDSS, produced as in Fig.~\ref{Fig1}. The lower
panel shows the vertical position of the centroid of the gas-layer 
as determined from the distribution of the \HI\ 
by Garc\'{\i}a-Ruiz et al., on the same radial scale and
with the aspect ratio restored to the real sky.}
\label{Fig8550}
\end{figure}

Table~\ref{Table3} shows that four systems (UGC 7089, UGC 7151, UGC 8709
and UGC 9242) have \HI-distributions that do not extend significantly
further than the optical image. Two of these (UGC 7089 and UGC 9242) show no
evidence for a truncation. These four systems are not suitable to investigate
relations between \HI-warps and disk truncations. These four systems do not
show clear evidence for \HI-warps anyway.

There is one system (UGC 8550) that shows no evidence for a truncation
nor for an \HI-warp. The information on this system is illustrated in
Fig.~\ref{Fig8550}; the radial extent clearly keeps growing at
fainter levels. The \HI\ does extend well beyond the optical
radius, but there is no convincing evidence for a warp. 

\begin{figure}
\centering
\includegraphics[width=8.8cm]{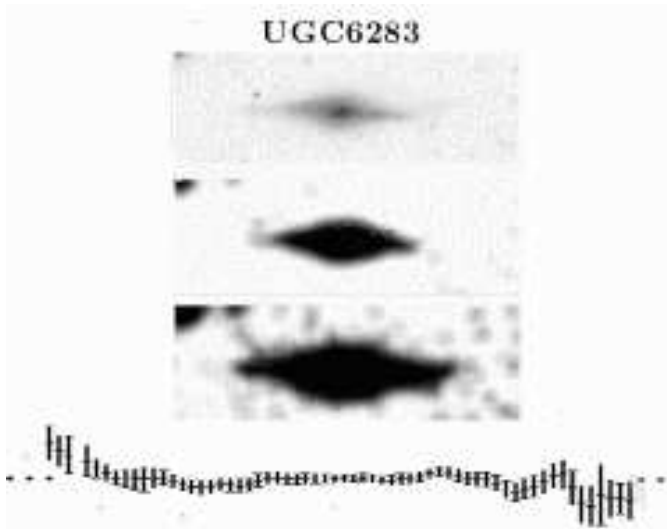}
\caption{UGC 6283 shows no evidence for a truncation, but it has 
a warped \HI-layer that sets in (projected onto the sky) at a radius 
larger than the optical radius. Further as in Fig.~\ref{Fig8550}.}
\label{Fig6283}
\end{figure}

Three galaxies have no truncation, but do possess warped \HI-layers,
namely UGC 6283, UGC 7090 and UGC 8396. The first has been illustrated
in Fig.~\ref{Fig6283}. The warp starts at about 3 arcmin from the center,
while the optical disk can be traced to 2.1 arcmin. It is of course
possible that this galaxy has a truncation at a
fainter level more closely to where we see the onset of the
\HI-warp. In the other two cases the onset of the warp on the sky starts
within the optical extent. The value for $R_{warp}/R_{deep}$ is 0.6 -- 0.7 
for both UGC 7090 and UGC 8396. 

There are three  galaxies (UGC 4278, UGC 4806 and UGC 7283) that 
show a truncation but have not unambiguous evidence for a
warp even though the \HI\ extends beyond the truncation 
radius. 

\begin{figure}
\centering
\includegraphics[width=8.8cm]{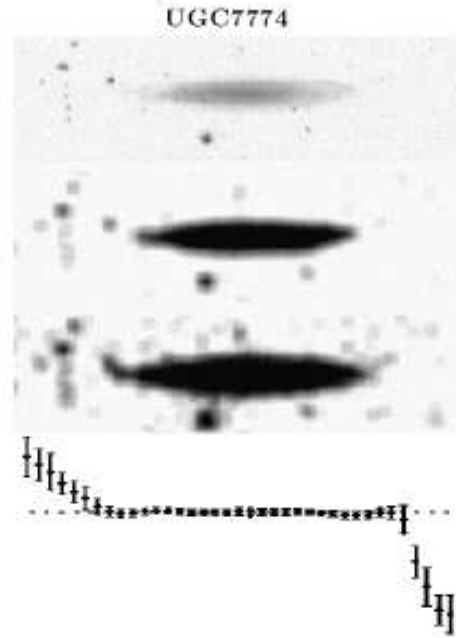}
\caption{UGC 7774 has a truncation and a warped \HI-layer that sets in 
(projected onto the sky) at about the truncation radius. 
Further as in Fig.~\ref{Fig8550}.}
\label{Fig7774}
\end{figure}

\begin{figure}
\centering
\includegraphics[width=8.8cm]{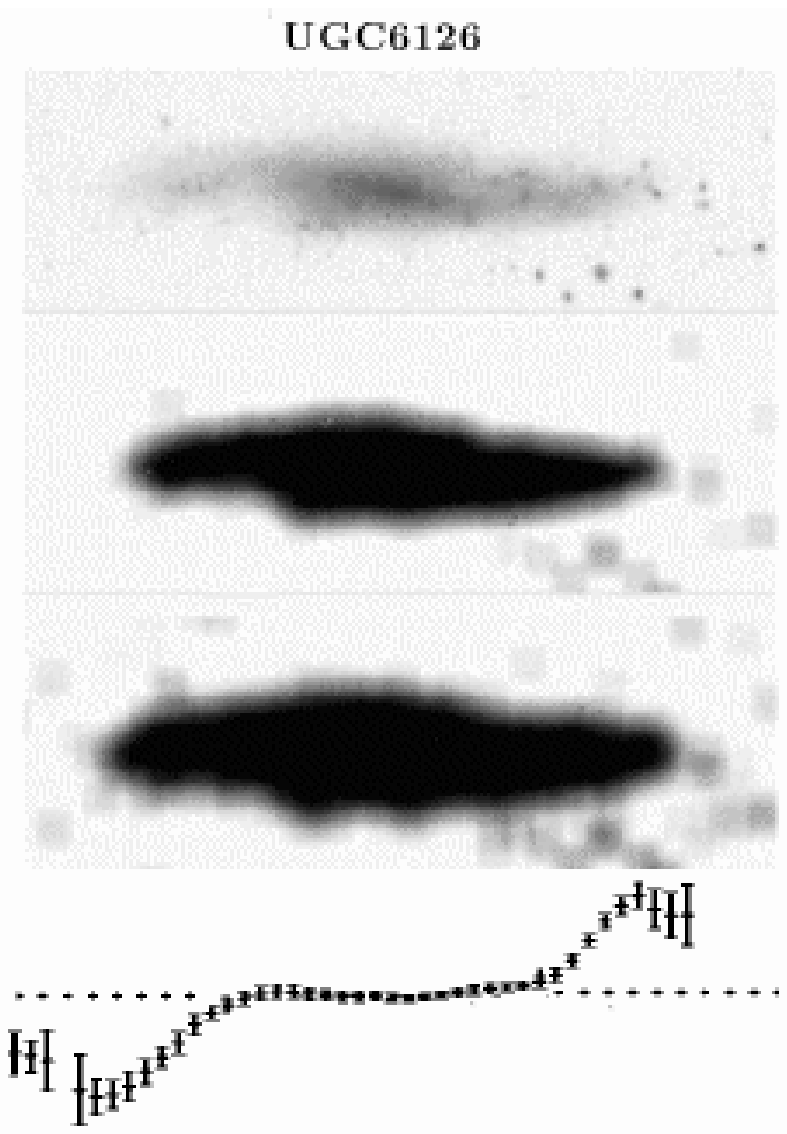}
\caption{UGC 6126 has a truncation and a warped \HI-layer that sets in 
(projected onto the sky) well within the truncation radius. 
Further as in Fig.~\ref{Fig8550}.}
\label{Fig6126}
\end{figure}

So, we are left with 12 galaxies with both a truncation and a warp. The
distribution of the ratio of the truncation radius and that where on the
sky the warps starts ($R_{warp}$)
is given in Table~\ref{Table4}. These include all the
``bona fide warped'' galaxies of Garc\'{\i}a-Ruiz et al. except UGC 7125, 
which I do not consider, and none of their ``non-warped'' galaxies.
I illustrate two cases in 
Fig.~\ref{Fig7774} and Fig.~\ref{Fig6126}. In UGC 7774 the
warp starts just beyond the truncation radius and for UGC 6126
well within this radius. 

The projection of a warp onto the plane of the sky is of course such
that the apparent onset is less than in the plane of the galaxy,
except when the direction of maximum warp
is perpendicular to the line of sight. 
In this case the warp is most readily recognised. If
we designate that configuration with an angle $\phi $ equal to
90$^{\circ }$, then other situations will occur in a fraction proportional to 
cos $\phi $, if
there is no effect of the orientation on the detectability of the warp.
This is obviously not the case; when the line of sight coincides with 
the line of
nodes, we expect to see a general thickening of the disk with faint
outer parts in the vertical profiles of the \HI-distribution, but not a
deviation of the plane of symmetry. The three galaxies with truncations but
no evidence for an \HI-warp 
(UGC 4278, UGC 4806 and UGC 7283) could be examples where 
the angle between the line of nodes and the line of sight are relatively 
small. These are then probably small warps, possibly involving little 
\HI\footnote{The \HI\ extends only 10 to 20\%\ beyond $R_{max}$ on the
sky.}, as these warps would otherwise have been 
noticable in the channel maps. Add to this
UGC 8550 with no warp or truncation. Also in two of the three galaxies with
a warp but no truncation (UGC 7090 and UGC 8396)
the onset of the warp is well within the optical 
extent and $R_{warp}$ is less than 0.7  $R_{deep}$. So we have at least six 
galaxies where a warp is possible at a projected $R_{warp}$ less than 0.6
$R_{max}$.

I now assume that the ratio $R_{warp}/R_{max}$ in the intrinsic
geometry is constant between all galaxies
and then calculate the resulting statistical distribution on the sky.
In Table~\ref{Table5} the observed and expected distributions have been 
compared. The expected distributions are for three values of
$R_{warp}/R_{max}$ in the plane of the galaxy. 
The value in the last row shows the probability that
the observed and expected numbers of galaxies have been drawn from the
same distribution according to a Kolmogorov-Smirnov test. 
The conclusion is that the observations 
are in good  agreement with the statement
that if a galaxy has an \HI-warps, it starts at a radius of 1.1 $R_{max}$.

\section{Relation with \HI-distributions and rotation curves and comparison to
other samples.}

First I stress that disks within their optical radius are, except for the
very minor warps in the stellar disks close to the edges,  extremely flat. 
The deviations from a straight line in dustlanes in perfectly edge-on
systems such as NGC 7814 or NGC 4013 are less than one percent. In \HI\
studies of spiral galaxies that are very close to face-on (see van der
Kruit \&\ Shostak,  1982, 1984; Shostak \&\ van der Kruit, 1984) the
residual velocity field after the subtraction of a symmetric rotation
field has a random pattern with an average amplitude of only a few
km sec$^{-1}$. This shows that systematic vertical motions of the \HI\ are 
extremely small and indicates that \HI-disks, and therefore the stellar disks
in whose gravitational field the gas moves, must be extremely flat,
within the boundaries of the stellar disks. 

It can also be
illustrated by examining the edge-on galaxies studied here.
In the bottom panels of Fig.~\ref{Fig8550} to
Fig.~\ref{Fig6126} I have restored the aspect ratio in order to show the
deviations of the \HI-layer with the same scale in the radial and the
vertical directions. Within the
optical extent the \HI\ ridge-lines are very flat; I note that 
UGC 6126 may not be a good example, since it is
not entirely edge-on and shows in the optical picture spiral structure. 
In particular UGC 7774 shows a very
flat disk which suddenly starts to warp.
This quite {\it discontinuous} and {\it abrupt} change from very flat to
significantly and sometimes very strongly deviating from the inner plane is
a feature that is often observed 
and we have seen here that this occurs  close to the truncation
radius of the stellar disk. 

The abruptness of the onset of the warps and the close radial coincidence
with the truncation radius of the stellar
disk suggest that the \HI\ in the disk and that in the warped outer part might
have a distinct origin. In order to investigate this further I
will look at possible features in the \HI-distributions and the rotation
curves at the positions where the warps start.

Of the eight prominent edge-on galaxies in the northern hemisphere van der
Kruit \&\ Searle sample, all have disk truncations, but not all have \HI\ 
beyond that radius or pronounced warps. 
 For this one needs samples
that have both surface photometry and \HI-data and that is available
in  the van der
Kruit \&\ Searle sample of edge-on galaxies. I will discuss them in turn. 

NGC 891 was observed by Sancisi \&\ Allen (1979), 
Rupen (1991 and Swaters et al. (1997). 
No clear evidence for a warp is seen, but on the SW
side there is \HI\ beyond the stellar disk and there appears a feature in both
the \HI-distribution and the rotation curve at about the position of the
disk truncation (see Fig.s 
18-20 in Rupen or Fig.~2 in Swaters et al.).

The \HI-warp in NGC 4013 was discoverd by Bottema et al. (1987) and studied
by Bottema (1995, 1996). It is a extensive and very symmetric warp, starting 
at about the truncation radius. The onset of the warp is accompanied
by a sudden drop in the rotation curve and the deconvolved
\HI\ surface density shows a steep decrease to a much lower level there. NGC
4013 is a prime example to show a very flat inner
disk that suddenly and abruptly starts to warp near the truncation radius.

\HI\ observations of NGC 4217 have been reported in Verheijen (1997) and
Verheijen \&\ Sancisi (2001). The \HI\ does not extend significantly 
beyond the truncation radius.

NGC 4244 is a late-type pure-disk galaxy. The \HI\ observations of 
Olling (1996) show a slight warp in the outer parts, which sets in at
about the edge of the stellar disk. At this position the rotation curve
(Olling's Fig.~7 or 9) shows the onset of a decline; the \HI\ surface
density shows a sharp decrease at this radius (Olling's Fig 6, lower
panel). 

\begin{table}
\caption{Galaxies that have both a truncation in the stellar disk and a
warp in the \HI-layer and the parameter $R_{warp}/R_{max}$.}
\label{Table4}      
\centering           
\begin{tabular}{c l}    
\hline\hline                 
$R_{warp}/R_{max}$ & UGC \\    
\hline                     
1.2 & 3137 \\
1.1 & 5452, 7774 \\
1.0 & 2459, 8286 \\
0.9 & 3909, 5459, 7321 \\
0.8 & 6126, 6964, \\
0.7 & 1281, 8711 \\
\hline                                  
\end{tabular}
\end{table}

NGC 4565 has both an \HI-warp and a truncation in the stellar disk.
On the sky the onset of the warp occurs close to the
truncation in the stellar disk (e.g. Rupen, 1991). Fig.~27 in Rupen's
paper shows the ``extreme velocities'' channel maps and this demonstrates
the point that the inner disk is very flat and that the warp sets in
suddenly and abruptly. There is no detailed analysis of these data in
terms of a rotation curve or a radial \HI-distribution. Rupen's $lv$ plots
(particularly his  Fig.~16) show signs of drops in \HI\ surface density 
and rotation velocity near the onset of the warp, but this needs detailed
analysis (but see his Fig.~22 with the total \HI. 
The stellar disk of NGC 4565 also has a moderate warp (van der Kruit \&\ 
Searle, 1981a; see also Wu et al., 2002) in the same direction as in the \HI. 

NGC 5023 is a dwarf galaxy with the weakest evidence for a truncation in
the stellar disk among these galaxies. 
The \HI-observations of Bottema et al. (1986) show
that the \HI\ extends somewhat beyond the optical disk, but there is no 
warp and no clear features in the rotation curve or in the surface density
distribution have been observed.

The classical example of an edge-on warp
is in NGC 5907 (Sancisi, 1976). Casertano (1983) has shown that there is
a drop in the rotation curve just at the truncation radius 
and he relates this to a sharp drop in
density in the stellar disk. The radial distribution of the \HI\ has
been part of Casertano's modelling; it appears from his figures to be
dropping, but behaves relatively smoothly at the transition between the 
inner disk and the warp at about the truncation radius. 

\begin{table}
\caption{Distribution of the parameter $R_{warp}/R_{max}$ observed
and expected from random orientations. The bottom line shows the
probability that the observed and expected numbers are drawn from the
same distribution.}
\label{Table5}      
\centering           
\begin{tabular}{c c c c c}    
\hline 
\underline{$R_{warp}$} & 
Observed & $R_{warp}=$ & $R_{warp}=$ & $R_{warp}=$  \\     
$R_{max}\ \ \ $ & & $1.0 R_{max}$ & $1.1 R_{max}$ & $1.2 R_{max}$  \\    
\hline                     
1.2 & 1 & -- & -- & 3.3 \\
1.1 & 2 & -- & 3.5 & 2.5 \\
1.0 & 2 & 3.6 & 2.6 & 1.7 \\
0.9 & 3 & 2.7 & 1.8 & 1.4 \\
0.8 & 2 & 1.9 & 1.5 & 1.3 \\
0.7 & 2 & 1.6 & 1.4 & 1.2 \\
-- & 6 & 8.1 & 7.2 & 6.6 \\
\hline                                  
$p$ & & 0.706 & 0.963 & 0.538 \\
\hline
\end{tabular}
\end{table}

The early-type galaxy NGC 7814 
is very much spheroid-dominated. Yet, there is a straight dust-layer and a
relatively faint
\HI-disk. Recent deep Westerbrok observations (Kamphuis et al., 2006) have 
shown evidence for extra-planar \HI\ at faint levels. The geometry is still 
unclear. 

From this inventory it appears that a warp in the \HI\ generally starts
rather abruptly at the edge of the disk and is often accompanied by a
drop in the rotation curve and in the \HI\ surface density
distribution. The feature in the rotation curve 
would be related to the fact that the truncation radius corresponds not 
only to a rather sharp radial drop-off in the light
but also in the radial total density distribution. 

Next I look at those
systems in the Garc\'{\i}a-Ruiz et al. sample that have warps and appear
in Table~\ref{Table4}. 
Garc\'{\i}a-Ruiz et al. find that the radial surface density
profile of a {\it warped} galaxy has the typical form when normalised to
a fiducial $R_{\HI}$ (tabulated in Table~\ref{Table2}), 
which corresponds to a face-on surface density of 1.0
M$_{\odot }$ pc$^{-2}$ of \HI. The surface density is roughly
constant to slowly decreasing in the inner part, followed by a faster drop that
sets in around a 0.5 $R_{\HI}$, extending
out at low levels to 1.5 to 2 times $R_{\HI}$. They conclude then
(see also their Fig.~9) that {\it ``it is in
the second part of the profile (the steepest part, around the optical
edge) where a warp develops in most of the galaxies''}. Evidently 
it is a general feature that the \HI\ surface density drops steeply 
at the radius of the onset of
the warp. In Table~\ref{Table6} I look at possible features in the
rotaton curve as judged from Fig.~7 in Garc\'{\i}a-Ruiz et al. (2002).
Sometimes the rotation curve has not been derived into the warp region or is
significantly different on the two sides. In a number of
cases the rotation curve peaks near $R_{max}$ and/or shows a definite decline
just beyond that.

\begin{table}
\caption{Characteristics (from Fig.~7 in Garc\'{\i}a-Ruiz et al., 2002) 
in of the rotation curve at the
positions of the onset of the warps for the galaxies in Table~\ref{Table4}}
\label{Table6}      
\centering           
\begin{tabular}{c c l}    
\hline\hline
$R_{warp}\over R_{max}$ & UGC & \\    
\hline
1.2 & 3137 & Sharp drop in rotation curve at $R_{max}$ \\
1.1 & 5452 & No rotation curve given beyond $R_{max}$ \\
1.1 & 7774 & Drop in rotation curve beyond $R_{max}$ \\
1.0 & 2459 & No clear features near $R_{max}$ \\
1.0 & 8286 & Drop in rotation curve beyond $R_{max}$ \\
0.9 & 3909 & No feature in rotation curve near $R_{max}$ \\
0.9 & 5459 & Asymmetric; no consistent behavior near $R_{max}$ \\
0.9 & 7321 & Drop in rotation curve beyond $R_{max}$ \\
0.8 & 6126 & Drop in rotation curve on one side beyond $R_{max}$ \\
0.8 & 6964 & No obvious feature around $R_{max}$ \\
0.7 & 1281 & Asymmetric; no consistent behavior near $R_{max}$ \\
0.7 & 8711 & Rotation curve peaks near $R_{max}$ \\
\hline
\end{tabular}
\end{table}

For face-on samples we cannot easily measure the truncation radius.
However, following the rules of Briggs we may investigate the behaviour
of the surface density of \HI\ at the Holmberg radius. For definiteness
(see also the Appendix) I take the {\it face-on}
surface luminosity that corresponds to 26.5 B-mag arcsec$^{-2}$ for a
definition of a ``Holmberg'' radius. 
 Again one needs galaxies
that have both surface photometry and \HI-data and a useful sample then is  
the Wevers et al. (1986)
``Palomar-Westerbork Survey of Northern Hemisphere Galaxies''.
For each galaxy in this survey there is a listing
of the radial luminosity and \HI\ surface density profiles corrected to
face-on. In the B-band a surface luminosity of 1 L$_{\odot
}$ pc$^{-2}$ corresponds to 27.05 B-mag arcsec$^{-2}$, so with the
conversion to the J-band of Wevers et al. for a mean
(B-V) of 0.7 the ``Holmberg'' surface luminosity corresponds to 1.7 L$_{\odot
}$ pc$^{-2}$. I will designate this radius as $R_{Ho}^{fo}$.
For almost all  systems in the sample
$R_{Ho}^{fo}$ correlates with the steepest part
of the face-on \HI\ surface density profile and therefore is entirely in
agreement with the conclusion of Garc\'{\i}a-Ruiz et al. There are three
systems on which some more detailed mention should be made.

The first is NGC 628, which has a clear warp (Shostak \&\ van der Kruit, 
1984; Kamphuis \&\ Briggs, 1992), the plane of the \HI\ actually going
through the plane of the sky. Photometry in Shostak \&\ van der
Kruit possibly has revealed a truncation in the stellar disk. Both
occur at the same radius (see also discussion in van der Kruit, 2001)
and it also corresponds to $R_{Ho}^{fo}$.
The radial \HI-profile drops steeply at the radius of the onset of the
warp and then 
continues with a much slower decline. A similar picture emerges for
NGC 3726, for which the velocity field shows very clear signs of a kinematical
warp, that starts at about 3 arcmin radius; this is close to $R_{Ho}^{fo}$ and 
again beyond the onset of the warp the \HI\ surface density profile flattens. 
In NGC 5055 $R_{Ho}^{fo}$ comes out as 8.1 arcmin (or
17 kpc on the distance scale of Battaglia et al., 2006). The radius where
the tilted ring model of Battaglia et al. starts to show the onset of
the warp is somewhat further in (more like 11 or 12 kpc). The
luminosity profile of Wevers et al. shows a well-defined exponential
disk out to 11 arcmin (23 kpc; the observed surface brightness is
27.5 J-mag arcsec$^{-2}$ there), well into the \HI-warp. The radial \HI\
surface density profile flattens off and becomes almost flat just beyond
$R_{Ho}^{fo}$. 

After examining all galaxies in the Wevers et al. sample
 it appears to be the case that near $R_{Ho}^{fo}$ there generally is a
steep gradient in the \HI\ surface density and that when there is a warp 
(judged from the velocity field)
this usually develops in this area, where also the
Holmberg radius would be located, in agreement with Brigg's rule. 
It remains to be
investigated to what extent the classical  Holmberg radius or
$R_{Ho}^{fo}$ corresponds to a truncation radius as we see it in 
edge-on galaxies. Alternatively, one should investigate the
relation between the truncation radii defined by Pohlen \&\ Trojillo (2006),
which occur at brighter levels of surface brightness than the
Holmberg radius, on the one hand and kinematical
evidence for warps in the gas on the other. Unfortunately, 
\HI-observations are not available for most galaxies in the
Pohlen \&\ Trujillo  
sample or difficult to obtain because of their angular size.

\section{Summary of warp properties and possible origins of warps and
truncations}

We then have the following general characteristics:\\
$\bullet $ Whenever a galaxy has an extended \HI-disk with respect to the 
optical disk, it has an \HI-warp (Garc\'{\i}a-Ruiz et al.).\\
$\bullet $ Many galaxies, but not all,  have relatively sharp truncations, 
which are most easily identified in edge-on systems (Kregel et al., 2002; 
this paper). Evidence for truncations or breaks in the radial surface
brightness profiles has been found in less inclined systems (see Pohlen \&\ 
Trujillo, 2006), but that requires much more carefull photometry then 
the method employed here.\\
$\bullet $ When an edge-on  galaxy has an \HI-warp the onset occurs
just beyond the truncation radius (this paper). In less inclined
galaxies the warp is seen at the boundaries of the observable optical
disk, which corresponds to the Holmberg radius (Briggs, 1990.).\\
$\bullet $ In many cases the rotation curve shows a decline at about the
truncation radius (Casertano, 1983; Bottema, 1996; this paper), which
indicates that the truncation does not only occur in the light, but also
in the mass.\\
$\bullet $ The onset of the warp is abrupt and discontinuous (see discussion
above) and coincides in the large majority of
cases with a steep slope in the radial \HI\ surface density distribution
(Garc\'{\i}a-Ruiz et al.; this paper) after which this distribution flattens
off considerably. In galaxies, where the \HI\ does not extend beyond the 
optical radius the decline is of course even steeper, but the point here is 
that the character of the \HI-layer changes substantially at the radius where
the warp starts (see also Briggs, 1990).\\
$\bullet $ The inner disks are extremely flat in the dust and gas
distributions (see discussion above), while stellar disks show only very small
warps, if any (de Grijs, 1997). The onset of the warp is {\it abrupt} (this
paper) and beyond that the plane of the warp defines a ``new
reference frame'' (Briggs, 1990).

These characteristics seem to point to a picture in which the inner disk
out to the boundaries of the stellar disk on the one hand and the outer warped
\HI-layers on the other are distict components. 
The ``new reference frame'' that Briggs (1990) quotes manifests itself in
inclined galaxies as a change in the line-of-nodes inside
and outside the transition radius where the warp starts. Briggs
discusses this in the context of the model for bending modes in
self-gravitating disks of Sparke \&\ Casertano (1988). 
These authors do find that discrete bending modes are possible, starting at 
about the truncation radius. However, they also conclude that these are
only discrete when the truncations in the stars and in the gas are very
sharp, occurring in a fraction of a scalelength. This is
much sharper than is observed. Models with oblate dark matter
halos allow discrete modes independent of the details of the truncation
itself. However, as pointed out by Briggs (1990), these models require
that the disks inside and outside of the onset of the warp should have a
common line-of-nodes, contrary to what is observed. On this basis Briggs
concludes that the lack of a common line-of-nodes throughout the entire 
warped disk predicted by models that rely on
normal bending modes to maintain warp coherence at all radii argues
against such models.

Briggs (1990) already concluded that these are two different
``regimes'', each having its own reference frame and states that the
physical significance of this is not clear. 
I stress here the abruptness of the transition and the
coincidence of the truncation radius
with the sharp drop in \HI\ surface density. This fits 
in a picture in which the two components are formed during separate epochs.
The first, which may last for a number of Gyr, is the
formation of the inner disk. It is not important for this
discussion whether that is monolithic or proceeds in a hierarchical manner.
Disk then have natural boundaries associated with the maximum specific
angular momentum of the material from which it formed.
Stellar disks then are well-bounded
structures with little \HI\ beyond their boundaries when this phase is
completed. The orientation in space
results from the integrated angular momentum that has been
collected in it. 

The second phase comes at a later stage, when material falls in in
the form of gas with generally a different orientation of angular
momentum. A major part will have higher specific angular
momentum then that at the truncation of the disk and will settle at
larger radii and in a different plane. The
fundamental point is the dichotomy between the
inner disk and the warped outer \HI-layer. The outer reference frame of
Briggs would then correspond to the orientation of the angular momentum
of the material that settles in the warped outer disk.

In this picture the truncations in the stellar disks would result from
the maximum specific angular momentum present 
in the disk after the first epoch of
formation of the inner disk. This would mean also that the origin of these
truncations is not related to a threshold for the occurence of star formation,
since there would be little gas beyond the truncation to form stars. 
Galaxies that look like this still exist; 
in the Garc\'{\i}a-Ruiz et al. sample
there are four galaxies with little or no \HI\ beyond the optical extent.
E.g. UGC 7151 does show a truncation, but no gas beyond that radius. For
such cases there is no need to invoke an inhibition of star formation
since there is essentially no gas in the first place.

The observation that the central surface brightness appears to correlate
with the ratio $R_{max}/h$, indicates that the surface brightness at the
radius where the
truncation occurs is roughly constant between galaxies (Kregel \&\ van der
Kruit, 2004; Pohlen \&\ Trujillo, 2006) and this in turn suggests
that this is also the case for the mass surface densities. 
This fits in the picture of a
threshold for star formation, which would imply that there is only a truncation
in the light, but not in the mass. The most likely instability  to 
produce this is gravitational instability as described by 
the Goldreich--Lynden-Bell (1965) criterion. As van der Kruit \&\ 
Searle (1982a) already showed, this criterion appears to be a 
poor predictor of the truncation radius, although it may be adapted
as by Schaye (2006) to better fit the data. More importantly, 
a number of detailed analyses
(in particular in NGC 5907 and NGC 4013; see above) indicate that the
rotation curves show the effects of a truncation in the mass, as would be
expected in the model where the truncation results from a maximum in
the specific angular momentum.  
The correlation of the ratio $R_{max}/h$ with the face-on surface brightness 
(righthand panel in 
Fig.~1 of Kregel \&\ van der Kruit) then  indicates that some
redistribution of angular momentum is required in this picture.

In most cases the truncation radius seems to correspond to very
steep slopes in the current gas surface density. The way this
fits into the picture suggested here is the following. After formation
of the inner disk, there was presumably little or no gas left
beyond the truncation radius and the
gas we observe today fell in later over a protracted period. Then there
is no need to invoke a threshold for star formation as the origin
of the truncations. At the end of the formation of the stellar disk
there simply was no gas present at radii beyond the truncation.

The picture, where the truncation is explained as a result of a collapse 
of a protogalaxy with a well-defined maximum in the specific angular
momentum distribution (van der Kruit, 1987), the protogalaxy is
assumed to be a Mestel sphere (with uniform density and uniform
rotation). In the case
of detailed conservation of angular momentum during a collapse in the
potential field of a dark halo that corresponds to a flat rotation curve, 
the resulting disk then becomes automatically  exponential as observed
while the truncation radius occurs around 4.5 scalelengths.
Observations suggest that this ratio is more like 4.0 $\pm$\ 1.1
with a tendency to increase for smaller values of the
scalelength (Fig.~1, left panel in Kregel \&\ van der Kruit, 2004; 
see also references
to earlier work therein), suggesting that the initial distribution 
is not precisely like a Mestel sphere and/or that there is
no complete conservation of specific angular momentum.

The referee of this paper has asked me to consider alternatives to this 
picture of late accretion events for the gas in the warp. In particular in the 
hierarchical model, one would expect that many accretion events (of small
galaxies and gas clouds) would occur and then it is not obviouw why the 
warped outer plane would represent a well-defined, single reference frame
(one of Brigg's rules of behaviour). So the alternative picture in which 
the gas was already present when the warp was 
formed as result of environmental effects, such as for example 
through tidal interactions (Reshetnikov \$\ Combes, 1998) 
or as a result of a galaxy disk moving through a low density intergalactic
medium (e.g. L\'opez-Corredoira et. al, 2002) could offer explanations as 
well.

I think these explanantions do work as well.
The important point remains that
the very flat inner disk should not be affected by these processes and this
can be accomplished if indeed these inner disks form relatively quickly in
the evolution of the galaxy. They would have to be massive enough to 
evolve into a stable, very flat disk with a
well-defined outer boundary that corresponds to the currently observable 
truncation radius. If then subsequently \HI\ is accreted 
with higher angular momentum on a 
somewhat longer timescale that comes in with a range of orientations of 
angular momentum then the processes of tidal interaction or of moving through 
an intergalactic medium could be responsible for rearranging this into
well-defined warps with a single reference frame. 
This is of course rather speculative, 
but the basis point remains that there are
two distinct epochs, one in which the inner disks forms and a later
one where the outer material is accreted and subsequently rearanged into
the observed warps in the \HI.

The referee also asked my to comment on why the warp starts at 1.1 
$R_{max}$ and not some other value if material with various different
angular momenta is accreted. On these I have to be even more speculative.
If indeed the first phase gives rise to a very flat stellar disk, then
this should be a fairly massive and rigid structure that conserves its
flatness very well, except maybe very close to the edges, where the
(minor compared to \HI) stellar warps are found. One then has to assume that
subsequently
material is accreted with a range of angular momenta and that material that 
would have specific angular momenta corresponding to radii within the
disk boundaries settle in it without much disturbing the flatness of
the stellar disks. This is maybe not unlikely in view of the much larger
amount of mass already collected in the inner disks during the earlier
phase. Material with higher specific angular momenta settle at a range of 
radii, from just outside the disk to at much larger distances. Again we will 
have to assume that the inner disk is then sufficiently rigid so that it is not
significantly disturbed when this outer material is rearranged into the 
regular warp with a well-defined ``reference frame''. It is not unreasonable 
to assume that such a disk looses the capability to attract gas towards
its plane --as it does when it is inside its outer edge-- not much beyond
its boundary at the truncation radius $R_{max}$ and a mean value of 1.1 
$R_{max}$ does not seem surprising.

\section{Conclusions}

\begin{enumerate}
\item The Sloan Digital Sky Survey (SDSS) can be used in a simple way to
investigate the presence of truncations in the disk light distributions
in a uniform and consistent manner.
\item The edge-on galaxies in the sample of Garc\'{\i}a-Ruiz et al.
(2002) shows in the majority (17 out of 23)
of cases evidence for truncations in their stellar disks.
\item When an \HI-warp is present it starts at about 1.1 $R_{max}$ (the 
truncation radius).
\item There is a discontinuous and abrupt transition to the warped disk,
occuring at the truncation of the stellar disk, where also the \HI\
surface density drops steeply.
\item These findings suggest that the inner flat disk and the outer
warped disk are distinct components with a quite separate formation
history, probably during quite different epochs.
\item This picture is consistent with the view that truncations result from
the maximum specific angular momentum in the material that formed the inner
disk and there is no need to invoke a threshold to star formation to
explain the occurence of truncations in edge-on galaxies.
   \end{enumerate}

\begin{acknowledgements}
A significant fraction  of this work was done during a short sabattical
visit to the Space Telescope Science Institute in Baltimore. I thank the
director, Matt Mountain, for financial support and Ron Allen for
organizing this visit and stimulating discussions with him and 
his students and postdocs. The visit was also
supported from the special grant that the Faculty of Mathematics and
Natural Sciences of Groningen University puts at my disposal in
connection with my Jacobus C. Kapteyn honorary professorship.
The manuscript was revised during the stay as a visiting scientist
at the Santiago offices of the European Southern Observatory;
I am grateful to Catherine Cesarzky and Felix Mirabel for hospitality
during this period.\\
I thank Koen Kuijken and Renzo Sancisi for comments on 
drafts of this paper and for permission
to use figures from the Garc\'{\i}a-Ruiz et al. paper. I thank
the referee for carefully reading the manuscript and making detailed 
comments and suggestions.\\
This research made use if the Sloan Digital Sky Survey.
unding for the SDSS
has been provided by the Alfred P. Sloan Foundation, 
the National Aeronautics and Space 
Administration, the National Science Foundation, the U.S. Department of 
Energy, the Japanese Monbukagakusho, and the Max Planck Society.\\
Use was also made of the Digitized Sky Surveys.
The DSS were produced at the Space Telescope 
Science Institute under U.S. Government grant NAG W-2166. The images of 
these surveys are based on photographic data obtained using the Oschin 
Schmidt Telescope on Palomar Mountain< california and the UK Schmidt 
Telescope on Siding Spring Observatory, Australia. 
\end{acknowledgements}

\appendix

\section{The Holmberg radius and Holmberg's observations of edge-on galaxies}

The Holmberg radius (and for that matter $R_{25}$; the radius
at 25 B-mag arcsec$^{-2}$ in the RC3: de Vaucouleurs et al.,
1991) refers to levels of surface brightness that are {\it not}
corrected for the geometrical effects of the inclined orientation of the disk. 
In principle this 
correction should be performed in comparative studies such as 
that of Briggs (1990) and, although not entirely straightforward 
when no detailed surface photometry is available, can be done
reasonably reliably in a statistical sense. The geometrical correction is not
small and the effect of applying it is not negligible. For
example, for a face-on disk with a central surface brightness of 21.65 B-mag
arcsec$^{-2}$ (Freeman, 1970) $R_{25}$ is at 3.09$h$ ($h$ is the exponential
scalelength) and $R_{26.5}$ at 4.47$h$. 
If the galaxy is inclined by 45$^{\circ }$ the observed central surface 
brightness has brightened to 21.27 B-mag arcsec$^{-2}$ and we find then 
$R_{25}$ at 3.43$h$ and $R_{26.5}$ at 4.81$h$ along the major axis. 
These are differences of the order of 10\%\ and a substantial
fraction of Briggs' galaxies are more inclined then 45$^{\circ }$. I describe
and discuss here the definition of the Holmberg radius and how
Holmberg derived them.

It appears that there is much ignorance and confusion
about the origin of the concept of the Holmberg radius and how it
was originally defined. For example, 
the first four items that I found after doing a search for ``Holmberg Radius''
in Google gave the following links and descriptions:\\
www.daviddarling.info/encyclopedia/H/Holmberg\_radius.html:\\
{\it ``A criterion, developed by Erik Holmberg in 1958, to estimate the size of 
a galaxy without regard to its orientation in space. It is the radius at 
which the surface brightness is 26.5 magnitudes per square arcsecond in blue 
light.''}\\
scienceworld.wolfram.com/astronomy/HolmbergRadius.html:\\
{\it ``The radius of the isophote of an elliptical galaxy corresponding to a 
surface brightness of 26.5 blue magnitudes per square arcsecond.''}\\
www.site.uottawa.ca:4321/astronomy/index.html\#Holmberg-\linebreak[4]
radius:\\
{\it ``The radius of an external galaxy at which the surface brightness is 
26.5 mag arcsec$^{-2}$. This criterion was developed by Holmberg in 1958 
to estimate the actual dimensions of the major and minor axes of a galaxy 
without regard to its orientation in space.''}\\
de.wikipedia.org/wiki/Holmberg-Radius:\\
{\it ``Der Holmberg-Radius ist eine Methode zur Absch\"atzung des Radius einer 
elliptischen Galaxie. Er wurde nach dem schwedischen Astronom Erik Bertil 
Holmberg benannt.
Der Holmberg-Radius ist definiert als die L\"ange der gro\ss en Halbachse 
der Galaxie bis zu der Isophoten (...), die einen 
Bereich der Oberfl\"achenhelligkeit von $\mu_H $ = 26.5 mag/arcsec$^{2}$
umschlie\ss t. Gemessen wird dabei im sogenannten B-Band (...). 
Der Holmberg-Radius gibt also den 
Bereich einer Galaxie an, aus dem der gr\"o\ss te Teil des Lichtes kommt 
(26,5 Magnituden entsprechen dabei etwa 1 bis 2 Prozent der Helligkeit des 
Nachthimmels).''}\footnote{The Holmberg radius is a method to estimate the
radius of an elliptical galaxy. It was named after the Swedish astronomer
Erik Bertil Holmberg. The Holmberg radius is defined as the length of half
the major axis out to the isophote (...) that encloses
a surface brightness of $\mu_H $ = 26.5 mag/arcsec$^{2}$. 
This is measured
in the so-called B-Band (...). The Holmberg radius therefore
shows the extent of a galaxy from which the majority of the light
originates (26.5 magnitudes corresponds to about 1 to 2 percent of the
brightness of the night sky).} 
\\
There was no such entry in the English version of Wikipedia.

So, what is the definition of the Holmberg radius? The original reference is 
the study 
Holmberg published in 1958 entitled {\it ``A Photographic Photometry of
Extragalactic Nebulae; I. A Study of Integrated Magnitudes and Colors of
300 Galaxies''} (Holmberg, 1958), 
in which he reported on a heroic undertaking
of microphotometer tracings of photographic plates (both in the
photographic band with 103a-O emulsion and in the photovisual band with
103a-D or 103a-C emulsion behind a Schott 11 filter), mostly 
taken at Mount Wilson between 1947 and 1955. 
The 1205 individual magnitude measures were performed by scanning plates 
that on one half were exposed in focus to the galaxy
and the other half out of focus to a set of standard stars in the
North Polar Sequence or in Selected Area 57. The scans of the
galaxies were performed along the major and minor axes. This method goes
back to much earlier work by Holmberg (1937, 1946, 1950). 

Holmberg {\it defines} his diameters as the positions where the
relative photographic density (that of galaxy + sky compared to sky
alone) is 0.5\%. This, he notes, is {\it ``close to the practical measuring
limit''}, and {\it ``corresponds, on an average,  to 
surface magnitudes of 26\magspt 5  (photogr. reg.), and 
26\magspt 0 (photov. reg.)''} (Holmberg, 1958, p. 12). So, a Holmberg
radius is not defined at a specific surface brightness in a certain
photometric band and corrected for orientation, but rather
corresponds in a statistical sense among Holmberg's sample to certain
approximate observed surface
brightnesses in the photographic and photovisual bands. 
Subsequent to Holmberg's study, Holmberg radii have been
quoted as related to be the isophote at 26.5 magnitudes per square arcsec 
in the Johnson $B$-band, which is close to the photographic ({\it pg}) 
band\footnote{The transformation can be written as 
$B - m_{pg} = 0.2 - 0.1( B - V)$
(see Kormendy \&\ Bahcall, 1973, based on Arp, 1961).}. 
New and independent measurements of radii of galaxies appeared in the 
literature much later, since it took many years after Holmberg's pioneering 
study before surface photometry to the required level became available.

The notion that the Holmberg radius defines the dimension of a
galaxy ``without regard to its orientation in space'' is incorrect,
while it also is not true that it applies to elliptical galaxies.
So, why did Holmberg choose not to correct his diameters for the geometrical
effects of the orientation?
Holmberg (1958) discusses the effect of inclination on his
diameters on page 44: 
{\it`` However, the observed surface magnitude of a spiral system
depends on the inclination, and there thus is a possibility that the
diameter of a nebula having an edgewise orientation is measured too large
as compared to the diameter of a nebula oriented in the celestial plane.
According to the writer's opinion the systematic effect is probably
small. Firstly, the difference in surface magnitude between nebulae of
incl. 0$^{\circ }$ and 90$^{\circ }$ is of a moderate size on account of
the internal absorption, and, secondly, the surface magnitude gradients
are in the outmost parts of spiral nebulae rather steep.''} In Holmberg's
chapter in {\it Stars and Stellar Systems IX} (Holmberg, 1975;
but written before 1966) he estimates that {\it 
``the change in measured major diameter would be about 7 percent, if the
surface brightness is assumed to increase by a factor of 2''}. 

Holmberg asserted (largely on the basis of his statistical study of 
the apparent surface
brightness as a function of axis ratio on the sky) that the effects of
dust absorption would to a large extent compensate for the geometrical
effects such that no correction was required. 
There is now considerable evidence that in galaxy disks the opacity in
the outer parts, where the diameters are defined, is probably small
(Holwerda et al., 2005; see also Holwerda, 2005 and
Andredakis \&\ van der Kruit, 1992) and the
geometrical correction in the measurement of disk radii is
probably justified. 
It is of interest to do the comparison for the Wevers et al. (1986) sample,
where we can derive the radius at 26.5 J-mag arcsec$^{-2}$ (which is
very close to the B-band) $R_{26.5}$ and the radius at 1.7 L$_{\odot, J}$
pc$^{-2}$ face-on surface luminosity $R_{Ho}^{fo}$ and for most of which
there are original diameters derived by Holmberg (1958) $R_{Ho}$. The ratio
$R_{26.5}$ to  $R_{Ho}$ the is 0.98 $\pm$\ 0.12, which shows the remarkable
accuracy of Holmberg's photometry, especially considering the
observational difficulties that he faced and the --compared to current 
standards-- primitive techniques and equipment he had at his disposal. 
Comparison with $R_{Ho}^{fo}$
reveals of course the inclination distribution of the sample; in
this case the ratio $R_{Ho}$ to $R_{Ho}^{fo}$ is 1.22 $\pm$\ 0.16. So
there is statistically a substantial uncertainty depending on whether or
not application of the full geometric correction is justified or not.

The RC3 (de Vaucouleurs et al.,1991) lists $D_{\circ }$ as the
corrected (for extinction and inclination) diameter out to the isophote
of 25 B/mag arcsec$^{-2}$. This is derived from the uncorrected diameter
to that isophote $R_{25}$. Here there also is confusion. E.g.
Binney and Merrifield (1998, p.61) describe $D_{25}$ 
\underline{with
reference to the RC3} as {\it ``the diameter that one estimates
the $I = 25 \mu _{B}$ isophote would have if the galaxy were seen face on
and unobscured by dust''}. That is {\it not} $D_{25}$ in the RC3, which uses 
$D_{\circ }$ for the corrected diameter. 
Confusing the issue this way is bad practice. 

It is of interest to see what Holmberg quotes for the edge-on galaxies
in his sample. When the apparent sharp edges of galaxies such as NGC
4565 were first noted (van der Kruit, 1979), I felt it was so obvious
that edge-on disks as this galaxy do not grow in the radial direction
with deeper exposures that it would have been found before.
Holmberg's (1958) study was an obvious candidate. It
has six edge-on systems in its sample, of which 4 overlap with that
studied by van der Kruit \&\ Searle (1981a,b, 1982a,b), namely
NGC 891, 4244, 4565 and 5907. I analyzed the additonal two (NGC 4631 and 5746)
using the SDSS in the same manner as described in this paper. 
In NGC 4631 there was clear evidence for a truncation, but NGC 5746 shows
no obvious indication of such a feature. The measurements
are shown in Table~\ref{TableA}. 

\begin{table}
\caption{Dimensions (in arcmin unless ratios) of edge-on galaxies in 
Holmberg's (1958) sample}
\label{TableA}      
\centering           
\begin{tabular}{l c r c c c c}    
\hline\hline                 
Galaxy & $R_{max}$ & $R_{Ho}$ & ratio &  $(b/a)$ & $(b/a)_{\rm Ho}$ 
& ratio \\    
\hline                     
NGC \ \  891 & 7.0 & 7.5 & 1.07 & 0.25 & 0.29 & 1.16 \\
NGC 4244 & 8.8 & 9.0 & 1.13 & 0.16 & 0.14 & 0.88 \\
NGC 4565 & 8.3 & 10.0 & 1.20 & 0.18 & 0.17 & 0.94 \\
NGC 4631 & 8.5 & 9.5 & 1.12 & 0.23 & 0.25 & 1.09 \\
NGC 5746 & 4.6 & 4.5 & 0.98 & 0.27 & 0.23 & 0.85 \\
NGC 5907 & 5.8 & 7.8 & 1.34 & 0.13 & 0.16 & 1.23 \\
\hline                                  
\end{tabular}
\end{table}

For NGC 5746 the listed $R_{max}$
is the faintest contour that could be measured and which is indicated as
$R_{deep}$ in this paper. Remarkably, it is the only case where the
value listed by Holmberg is very close to the one derived fron the SDSS data. 
For the remaining galaxies
the Holmberg radii are larger than the truncation radii. 
The average ratio is 1.17 $\pm$\ 0.10. In particular for
NGC 4565 the radius that Holmberg finds is well beyond the radius where light
can be traced even on the deepest images available nowadays. 
Holmberg certainly did not see
the effect of a sharp decrease in surface brightness associated with the
truncations. However, his axis ratios are very comparable to the 
ones derived here; the average ratio is $1.02 \pm 0.16$. This implies that
Holmberg overestimated the extents of the minor axes also.

\end{document}